\documentclass[conference]{IEEEtran}
\usepackage{amsmath}
\usepackage{amsthm}
\usepackage{cite}
\usepackage{amssymb}
\usepackage{graphicx}
\usepackage{algorithm}
\usepackage{algorithmic}
\newtheorem{theorem}{Theorem}

\ifCLASSINFOpdf
\else
\fi

\begin{document}
\title{On the Characterization and Evaluation of Doppler Squint in Wideband ODDM Systems}
\author{
	Xuehan~Wang\textsuperscript{1},~Jinhong~Yuan\textsuperscript{2},~Jintao~Wang\textsuperscript{1},~Zhi~Sun\textsuperscript{1}\\
	\IEEEauthorblockA{
		\textsuperscript{1}Beijing National Research Center for Information Science and Technology (BNRist),\\
		Dept. of Electronic Engineering, Tsinghua University, Beijing 100084, China\\
		\textsuperscript{2}School of Electrical Engineering and Telecommunications, University of New South Wales, Sydney, NSW 2052, Australia\\
		wang-xh21@mails.tsinghua.edu.cn,~j.yuan@unsw.edu.au,~wangjintao@tsinghua.edu.cn,~zhisun@ieee.org}
}
\maketitle
\begin{abstract} 
The recently proposed orthogonal delay-Doppler division multiplexing (ODDM) modulation has been demonstrated to enjoy excellent reliability over doubly-dispersive channels. However, most of the prior analysis tends to ignore the interactive dispersion caused by the wideband property of ODDM signal, which possibly leads to performance degradation. To solve this problem, we investigate the input-output relation of ODDM systems considering the wideband effect, which is also known as the Doppler squint effect (DSE) in the literature. The extra delay-Doppler (DD) dispersion caused by the DSE is first explicitly explained by employing the time-variant frequency response of multipath channels. Its characterization is then derived for both reduced cyclic prefix (RCP) and zero padded (ZP)-based wideband ODDM systems, where the extra DD spread and more complicated power leakage outside the peak region are presented theoretically. Numerical results are finally provided to confirm the significance of DSE. The derivations in this paper are beneficial for developing accurate signal processing techniques in ODDM-based integrated sensing and communication systems.  
\end{abstract}

\IEEEpeerreviewmaketitle
\section{Introduction}
\label{sec_intro}
Future physical layer technologies are expected to enable ultra-reliable transmission over doubly-dispersive channels to support emerging high-mobility applications \cite{OTFS_tutorial,OTFS_sigmodel,RCP_ODDM,ref_ODDM}, such as intelligent transportation systems \cite{OTFS_tutorial}, non-terrestrial networks \cite{shi_NTN} and underwater acoustic (UWA) communications \cite{UWA_OFDM_DL}. To tackle the possible performance degradation of conventional orthogonal frequency division multiplexing (OFDM) systems over doubly dispersive channels, the orthogonal time frequency space (OTFS) and orthogonal delay-Doppler division multiplexing (ODDM) modulation have been proposed in \cite{ref_ODDM} to enhance the error performance by fully utilizing the time-frequency diversity. So far, OTFS and ODDM have been widely acknowledged as one of the waveform candidates for the next-generation wireless communication systems \cite{OTFS_tutorial,ODDM_tutorial}.\par 
Accurate channel estimation is critical for successful data detection in ODDM systems \cite{RCP_ODDM}, which highly relies on precise input-output (IO) characterization \cite{ODDM_tutorial}. The authors in \cite{ref_ODDM} first established the basic principle under DD domain orthogonal pulse (DDOP)-based implementation and on-grid channels. The processing is then simplified by utilizing the inverse fast Fourier transform (IFFT) and sample-wise pulse-shaping in \cite{ODDM_tutorial,ref_physical_IO_TCOM}. Based on the simplified framework, the IO relation was analyzed over general channels with off-grid delay time and Doppler shift \cite{ref_physical_IO_TCOM}.\par 
To achieve better coupling with DD domain sparse channel, DD resolutions of ODDM systems are required to be sufficiently high, which indicates large time and bandwidth occupation \cite{ODDM_tutorial}. As a result, the wideband effect should be considered to better characterize the time-varying channel \cite{DSE_THz} with interactive dispersion between the time-frequency domain and DD domain \cite{channel_time_orig}, which is also referred to as the Doppler squint effect (DSE) \cite{DSE_THz}. In \cite{OTFS_DSE_TWC_mine,OTFS_DSE_TVT_mine}, the IO relation was derived for OTFS systems considering DSE. Considering that OTFS modulation suffers from high out-of-band emission (OOBE) \cite{ref_ODDM,ref_physical_IO_TCOM}, it is necessary to investigate the characterization in the DD domain for more realistic ODDM waveforms.\par 
In this paper, we focus on the IO relation of ODDM systems with DSE. The phenomenon of interactive dispersion is first presented by analyzing the time-variant frequency response of wideband multipath linear time-variant (LTV) channels, which also explains the reason to refer to it as DSE. The time domain analysis is then established with DSE. Based on this, the IO characterization is developed for both reduced cyclic prefix (RCP) and zero padded (ZP) ODDM systems. Finally, numerical results are presented to confirm the significance of considering DSE.\par 
\section{System Model}
\label{sec_system}
In this paper, we consider a wideband ODDM system with $M$ multicarrier symbols and $N$ subcarriers within each multicarrier symbol, respectively. The overall system operates at the sampling frequency of $f_{s}=\frac{1}{T_{s}}$, where $T_{s}$ is also known as the delay resolution. The baseband signal model of the ODDM transceiver is first briefly reviewed. Then the multipath LTV channel model considering DSE is introduced to establish the basis of IO analysis. \par 
\subsection{ODDM Framework}
\label{subsec_ODDM}
In this paper, the approximate implementation of ODDM modulation is adopted by utilizing IFFT and sample-wise pulse-shaping like \cite{ODDM_tutorial,ref_physical_IO_TCOM}. Let $X[m,n]$ denote the data component at the $n$-th subcarrier within the $m$-th multicarrier symbol for $n=0,1,\cdots,N-1$ and $m=0,1,\cdots,M-1$. The normalized $N$-point IFFT\footnote{This operation is also widely known as the inverse discrete Zak transform.} is first utilized to transfer $X[m,n]$ to the $\dot{n}$-th sample within the $m$-th symbol as 
\begin{equation}
	x[m,\dot{n}]=\frac{1}{\sqrt{N}}\sum_{n=0}^{N-1}X[m,n]e^{j2\pi\frac{n\dot{n}}{N}}.
	\label{IFFT_ODDMTx}
\end{equation}
Considering the staggered property of ODDM modulation as shown in \cite{ref_ODDM,ODDM_tutorial}, the interval between adjacent multicarrier symbols is the delay resolution $T_{s}$ while the sample interval within each multicarrier symbol is $T=MT_{s}$ due to the upsampling. As a result, the baseband continuous waveform can be generated as
\begin{equation}
	s(t)=\sum_{m=0}^{M-1}\sum_{\dot{n}=-1}^{N-1}x[m,\dot{n}]a(t-mT_{s}-\dot{n}T),
	\label{Txbaseband_ODDM}
\end{equation}
where $a(t)$ denotes the truncated Nyquist pulse for the symbol interval of $T_{s}$. The duration\footnote{$2Q\ll M$ is required to guarantee the approximation precision.} of $a(t)$ is denoted as $2QT_{s}$, i.e., $a(t)=0$ holds for $|t|\geq QT_{s}$. Here, $x[m,-1]$ represents the prefix\footnote{In this paper, $M$ prefix samples are added to combat the equivalent delay spread less than $T$. The range can be adjusted if more accurate knowledge about the delay spread is known at the transmitter, however, it will not influence the IO relation because the sampling time at the receiver is fixed.} to combat the multipath spread. If RCP-ODDM system is considered like \cite{RCP_ODDM,ref_physical_IO_TCOM}, we have $x[m,-1]=x[m,N-1]$. If ZP-ODDM \cite{ZP_ODDM} is considered, the sum range for $\dot{n}$ can be set from $0$ to $N-1$ while the ZP range should be carefully investigated as illustrated in Section \ref{subsec_IO_ZP}.\par 
After appropriate receiver filtering, the received baseband continuous signal $r(t)$ is sampled at a period of $T_{s}$ to obtain time domain samples as $y[m,\dot{n}]=r(mT_{s}+\dot{n}T)$. The data component at the $n$-th subcarrier within the $m$-th symbol can then be derived by employing the normalized $N$-point fast Fourier transform (FFT) as
\begin{equation}
	Y[m,n]=\frac{1}{\sqrt{N}}\sum_{\dot{n}=0}^{N-1}y[m,\dot{n}]e^{-j2\pi\frac{n\dot{n}}{N}}.
	\label{FFT_receiver}
\end{equation}
\subsection{Wideband Multipath LTV Channel Model with DSE}
\label{subsec_channel}
After the up-conversion with the carrier frequency of $f_{c}$, the continuous-time passband signal $\tilde{s}(t)=\Re\left\{s(t)e^{j2\pi f_{c}t}\right\}$ is sent from the transmitter to the receiver via $P$ dominant paths. For ease of illustration, the impact of noise is temporarily disregarded like \cite{OTFS_sigmodel,OTFS_DSE_TWC_mine,ODDM_tutorial}. The received passband signal can be derived as \cite{channel_time_orig,OTFS_DSE_TWC_mine,OTFS_DSE_TVT_mine}
\begin{equation}
	\tilde{r}(t)=\sum_{i=1}^{P}\tilde{h}_{i}\tilde{s}\bigg(t-\left(\tau_{i}-\frac{\text{v}_{i}}{c}t\right)\bigg),
	\label{passband_continuous_IO}
\end{equation}
where $\tilde{h}_{i}$ and $\tau_{i}$ represent the attenuation and propagation delay associated with the $i$-th path, $\text{v}_{i}$ and $c$ denote the speed with which the $i$-th path length is decreasing and wave speed. After the down-conversion, the baseband received signal can be obtained as
\begin{equation}
	r(t)=\sum_{i=1}^{P}h_{i}e^{j2\pi\frac{\text{v}_{i}}{c}f_{c}t}s\bigg(t-\left(\tau_{i}-\frac{\text{v}_{i}}{c}t\right)\bigg),
	\label{baseband_continuous_IO}
\end{equation}
where $h_{i}=\tilde{h}_{i}e^{-j2\pi f_{c}\tau_{i}}$ denotes the baseband complex gain. In \eqref{baseband_continuous_IO}, if the time-variant delay $\frac{\text{v}_{i}}{c}t$ is much smaller than the delay resolution within a whole ODDM frame, it can be ignored to derive the DD domain approximation in \cite{ref_ODDM,ref_physical_IO_TCOM}. For ODDM systems, the delay resolution is $T_{s}$ while the frame duration can be approximated by $NMT_{s}$. As a result, the ratio between the maximum time-variant delay and the delay resolution can be derived as $\frac{\frac{\text{v}_{i}}{c}NMT_{s}}{T_{s}}=\frac{\text{v}_{i}}{c}NM$ like \cite{OTFS_DSE_TWC_mine}. Considering the typical value for $M=512$, $N=64$, and the maximum mobility of $1000$ km/h \cite{ref_ODDM}, the ratio is more than 3\%, which has a non-negligible effect on the system reliability. The significance of time-variant delay becomes larger considering the satellite communications \cite{OTFS_satellite} and UWA scenarios \cite{UWA_OFDM_DL,UWA_OTFS}. As a result, we investigate the IO characterization considering the wideband effect in \eqref{baseband_continuous_IO} and evaluate the model error by numerical experiments. To simplify the notation, let $b_{i}=\frac{\text{v}_{i}}{c}$ and $\nu_{i}=\frac{\text{v}_{i}}{c}f_{c}$ denote the Doppler scaling factor and Doppler shift at the carrier frequency, respectively.\par 
\begin{figure}
	\centering{\includegraphics[width=0.7\linewidth]{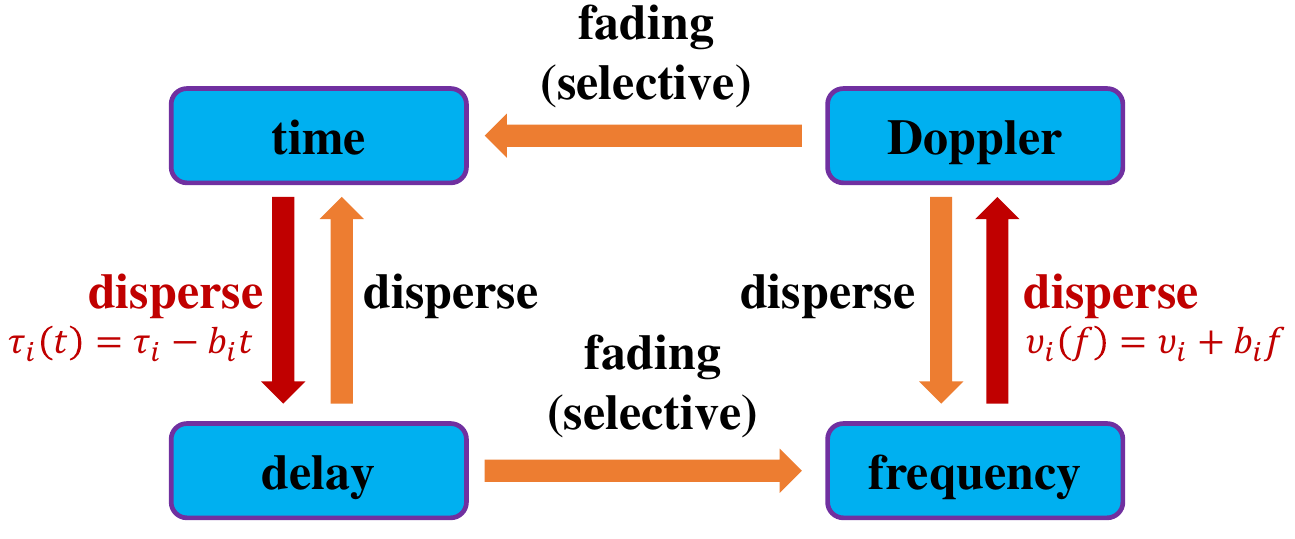}}
	\caption{The schematic of wideband multipath LTV channels considering the interactive dispersion.}
	\label{Fig_channel}	
\end{figure}
The wideband effect can be explained more clearly by considering the baseband time-variant frequency response $H(t,f)$, which can be derived \cite{channel_time_orig,OTFS_DSE_TWC_mine,OTFS_DSE_TVT_mine} as
\begin{equation}
	H(t,f)=\sum_{i=1}^{P}h_{i}e^{-j2\pi f\tau_{i}}e^{j2\pi\nu_{i}t}e^{j2\pi b_{i}ft},
	\label{TF_continuous}
\end{equation}
where $f$ denotes the baseband frequency. The difference\footnote{Please note that it does not mean the system is beyond the stationary time \cite{ref_physical_IO_TCOM} because the physical explanation remains the same.} beyond narrowband scenarios is the extra time-frequency coupled phase rotation as $e^{j2\pi b_{i}ft}$, which destroys the sparsity in the DD domain. As illustrated in Fig. \ref{Fig_channel}, the wideband effect leads to interactive dispersion. On one hand, the multipath spread causes the time-frequency dispersion, which is widely known as the doubly-dispersive channels. On the other hand, the time span leads to the time-varying delay as $\tau_{i}(t)=\tau_{i}-b_{i}t$ while the frequency span leads to the frequency-varying Doppler as $\nu_{i}(f)=\nu_{i}+b_{i}f$, which cannot be ignored. Due to the dual influences of interactive dispersion in the Doppler and delay dimension, we follow the prior literature in \cite{DSE_THz,OTFS_DSE_TWC_mine,OTFS_DSE_TVT_mine} to refer to this phenomenon as DSE.\par 
\section{Input-Output Characterization with DSE}
\label{sec_io}
In this section, the IO characterization is derived for ODDM systems considering DSE. The time domain analysis is first provided to serve as the basis. The relation is then established for RCP-ODDM systems. Then the appropriate range of ZP is investigated to make the characterization more simplified by avoiding inter-sample-interference (ISI). The impact of DSE is then discussed to clarify our contributions. For ease of illustration, we adopt the notations of normalized delay time and Doppler shift at the carrier frequency as $\tau_{i}=l_{i}T_{s}$ and $\nu_{i}=\frac{k_{i}}{NMT_{s}}$, respectively. The delay and Doppler spread are then respectively denoted as $l_{\min}\leq l_{i}\leq l_{\max}$ and $-k_{\max}\leq k_{i}\leq k_{\max}$. Without loss of generality, we set $l_{\min}$ and $l_{\max}$ as integers while $l_{i}$ and $k_{i}$ may not be integers.\par 
\subsection{Time Domain Analysis}
\label{subsec_IO_time}
In this subsection, we first investigate the time domain IO relation to serve as the basis for the following analysis. By substituting \eqref{Txbaseband_ODDM} in \eqref{baseband_continuous_IO}, the IO relationship between time domain samples can be derived as follows
\begin{equation}
	y[m,\dot{n}]=\sum_{m^{\prime}=0}^{M-1}\sum_{n^{\prime}=-1}^{N-1}x[m^{\prime},\dot{n}^{\prime}]\psi_{m,m^{\prime},\dot{n},\dot{n}^{\prime}},
	\label{time_io_ini}
\end{equation}
\begin{figure*}
	\begin{equation}
		\psi_{m,m^{\prime},\dot{n},\dot{n}^{\prime}}=\sum_{i=1}^{P}h_{i}\times\underbrace{e^{j2\pi\frac{k_{i}(m+\dot{n}M)}{NM}}a\Big(\big(m+\dot{n}M-m^{\prime}-\dot{n}^{\prime}M-l_{i}+b_{i}(m+\dot{n}M)\big)T_{s}\Big)}_{\psi^{i}_{m,m^{\prime},\dot{n},\dot{n}^{\prime}}}
		\label{time_base_ini}
	\end{equation}
	\hrulefill	
\end{figure*}where $\psi_{m,m^{\prime},\dot{n},\dot{n}^{\prime}}$ can be derived in \eqref{time_base_ini} at the top of the next page. Considering the limited time span of the Nyquist pulse $a(t)$, let $l=(m+\dot{n}M)-(m^{\prime}+\dot{n}^{\prime}M)$ denote the delay tap of the equivalent channel. Eq. \eqref{time_io_ini} can then be rewritten as
\begin{equation}
	y[m,\dot{n}]=\sum_{l=l_{\min}^{\prime}}^{l_{\max}^{\prime}}x[(m-l)_{M},\dot{n}^{\prime}]g_{l}(m,\dot{n}),
	\label{time_IO_tap}
\end{equation} 
where $\dot{n}^{\prime}$ is determined by $m-l$ and $\dot{n}$ while $l_{\min}^{\prime}$ and $l_{\max}^{\prime}$ denote the tap range of the equivalent sampled channel. The time-variant impulse response can be derived as $g_{l}(m,\dot{n})=\sum_{i=1}^{P}h_{i}g_{l}^{i}(m,\dot{n})$ with 
\begin{equation}
	\small
	\begin{aligned}
	g_{l}^{i}(m,\dot{n})&=e^{j2\pi\frac{k_{i}(m+\dot{n}M)}{NM}}a\Big(\big(l-l_{i}+b_{i}(m+\dot{n}M)\big)T_{s}\Big).
	\end{aligned}
	\label{time_tap_eachpath}	
\end{equation}
Eq. \eqref{time_tap_eachpath} reveals that both the Doppler shift and time-varying delay lead to the time-variant impulse response. As a result, the length of the sampled impulse is also time-varying rather than fixed like in \cite{ref_physical_IO_TCOM} while $l_{\min}^{\prime}$ and $l_{\max}^{\prime}$ represent the maximum tap range. Since we have $a(t)=0$ for $|t|\geq QT_{s}$, the tap range of the equivalent channel can be derived as
\begin{equation}
	\begin{cases}
		l_{\min}^{\prime}&=\lceil{l_{\min}-Q-b_{\max}(NM-1)}\rceil\\
		l_{\max}^{\prime}&=\lfloor{l_{\max}+Q+b_{\max}(NM-1)}\rfloor
	\end{cases},
	\label{tap_range}
\end{equation}
where $b_{\max}=\frac{v_{\max}}{c}$ denotes the maximum value of the Doppler scaling factor with $v_{\max}$ representing the maximum mobility. Without loss of generality, we set $l_{\min}^{\prime}=0$ by appropriate synchronization\footnote{If DSE is ignored, we have $l_{\min}=Q$. It is suitable due to the causal property of practical pulse-shaping configurations, where the group delay of $QT_{s}$ should be considered in the equivalent delay time.} as $l_{\min}=Q+\lfloor{b_{\max}(NM-1)}\rfloor$. Meanwhile, we assume that the maximum delay tap $l_{\max}^{\prime}$ is less than $M$ to simplify the analysis, which can be satisfied in most scenarios \cite{OTFS_DSE_TWC_mine,ref_physical_IO_TCOM,UWA_OTFS}. As a result, the value of $\dot{n}^{\prime}$ in \eqref{time_IO_tap} can be derived as
\begin{equation}
	\dot{n}^{\prime}=
	\begin{cases}
		\dot{n}, &m\geq{l}\\
		(\dot{n}-1)_{N}, &m<l
	\end{cases},
\end{equation}
which completes the time domain characterization.\par 
\subsection{RCP-ODDM Systems}
\label{subsec_IO_RCP}
Based on the analysis in the time domain, the following theorem can be derived to establish the IO characterization in the DD domain for RCP-ODDM systems.\par 
\begin{theorem}
	\label{th1_RCP_ODDM_IO}
	\rm
	The relation between $Y[m,n]$ and $X[m,n]$ can be formulated as
	\begin{equation}
		\label{RCP_ODDM_IO}
		\begin{aligned}
			Y[m,n]&=\sum_{l=0}^{l_{\max}^{\prime}}\sum_{k=0}^{N-1}X[(m-l)_{M},(n-k)_{N}]\\
			&\times \phi[m-l,n-k]G_{l}(m,k),
		\end{aligned}
	\end{equation}
	where $G_{l}(m,k)$ is presented in \eqref{RCP_DD_G} at the top of the next page. The phase rotation due to CP can be derived as
	\begin{equation}
		\label{DD_IO_extraPhase}
		\phi[m^{\prime},n^{\prime}]=
		\begin{cases}
			1, &0\leq m^{\prime}\leq M-1\\
			e^{-j2\pi \frac{n^{\prime}}{N}}, &-M+1\leq m^{\prime}<0
		\end{cases}.
	\end{equation}
	\begin{IEEEproof}
		The proof is provided in Appendix \ref{th1_RCP_ODDM_IO_proof}.
	\end{IEEEproof}
\end{theorem}\par 
\begin{figure*}
	\begin{equation}
		\label{RCP_DD_G}
		G_{l}(m,k)=\sum_{i=1}^{P}h_{i}\times\underbrace{e^{j2\pi\frac{k_{i}m}{NM}}\frac{1}{N}\sum_{\dot{n}=0}^{N-1}e^{-j2\pi\frac{\dot{n}(k-k_{i})}{N}}a\Big(\big(l-l_{i}+b_{i}(m+\dot{n}M)\big)T_{s}\Big)}_{G^{i}_{l}(m,k)}
	\end{equation}
	\hrulefill	
\end{figure*}
Compared with the narrowband analysis in \cite{ref_physical_IO_TCOM}, the differences in the DD domain IO characterization can be embodied in two aspects as follows.
\begin{itemize}
	\item [1)] \textbf{Extra delay-Doppler spread}: Since DSE leads to the time-variant delay associated with the mobility direction, the extra delay tap spread of $2b_{\max}(NM-1)$ will also be appended. In radio-frequency scenarios with the mobility of less than $1000$ km/h, the extra delay spread can be bounded by 2 as illustrated in \cite{OTFS_DSE_TWC_mine}. However, when considering more severe wideband effects like satellite and UWA scenarios, the impact will be more severe. Even though the tap range is time-variant, all possible equivalent taps will appear in the DD domain since full time response is collected. The dual effect can be embodied in the Doppler domain as well. In the meantime, the peak point for the channel response of each path in narrowband scenarios will be averaged into a peak region due to the time-variant delay and frequency-dependent Doppler.
	\item [2)] \textbf{More complicated power leakage outside the peak region}: It can be observed from \eqref{RCP_DD_G} that the pulse-shaping also changes the structure along subcarriers. As a result, the periodic sinc function cannot be derived along subcarriers like \cite{ref_physical_IO_TCOM}. It also leads to the delay-Doppler coupling in the channel matrix, which is dependent on the exact Nyquist pulse-shaping function. This indicates that the power leakage pattern outside the peak region will be more complicated. 
\end{itemize}\par 
Even though the closed-form characterization in the DD domain cannot be provided considering the complicated and various forms of Nyquist filters, the deduction in \textbf{Theorem \ref{th1_RCP_ODDM_IO}} can still serve as a qualified representation to evaluate the impact of DSE since \eqref{RCP_DD_G} has revealed the basic property of wideband effect. To mitigate DSE, the value of the ratio between the maximum time-variant delay and delay resolution as $b_{\max}NM$ should be sufficiently small, which indicates the trade-off between high DD resolutions and model accuracy. Based on the deduction in \textbf{Theorem \ref{th1_RCP_ODDM_IO}}, the matrix form of wideband characterization of ODDM input-output relation can also be established as
\begin{equation}
	\label{RCP_matrixform}
	\mathbf{y}_{\text{DD}}=\mathbf{H}_{\text{DD}}\mathbf{x}_{\text{DD}}+\mathbf{w}_{\text{DD}},
\end{equation}
where we have $\mathbf{y}_{\text{DD}},\mathbf{x}_{\text{DD}}\in\mathbb{C}^{NM\times1}$ with $\mathbf{y}_{\text{DD}}(mN+n)=Y[m,n]$ and $\mathbf{x}_{\text{DD}}(mN+n)=X[m,n]$. $\mathbf{w}_{\text{DD}}$ represents the additive noise samples in the DD domain while the DD domain channel matrix $\mathbf{H}_{\text{DD}}\in\mathbb{C}^{NM\times NM}$ can be derived as
\begin{equation}
	\label{RCP_channelmatrix}
	\begin{aligned}
		&\mathbf{H}_{\text{DD}}(mN+n,N(m-l)_{M}+(n-k)_{N})\\
		&=\begin{cases}
			G_{l}(m,k), &m\geq{l}\\
			e^{-j2\pi\frac{n-k}{N}}G_{l}(m,k), &m<l\\
		\end{cases}.
	\end{aligned}
\end{equation}
\subsection{ZP-ODDM Systems}
\label{subsec_IO_ZP}
The characterization in \textbf{Theorem \ref{th1_RCP_ODDM_IO}} reveals two distinct scenarios caused by the extra phase rotation $\phi[m-l,n-k]$ in \eqref{DD_IO_extraPhase}, which harms the low-complexity detectors \cite{ZP_ODDM,RCP_ODDM}. To tackle this challenge, a typical method is setting appropriate ZP symbols within the frame like \cite{ZP_ODDM,OTFS_ZP_MRCdetect_TVT}, i.e., $X[m,n]$ is set as zero for $m<m_{\min}$ and $m>m_{\max}$. It simplifies the IO relation by forcing the circular convolution along ODDM subcarriers, i.e., no ISI in the time domain. Similar techniques can also be extended to ODDM systems with DSE, which is illustrated in the following theorem.\par 
\begin{theorem}
	\label{th2_ZP_ODDM}
	\rm
	If the ZP range can be set as
	\begin{equation}
		\label{ZP_range_eq}
		\small
		\begin{cases}
			m_{\min}\geq\lceil{Q-l_{\min}-1+b_{\max}\big((N-1)M-1\big)}\rceil\\
			m_{\max}\leq\lfloor{M-Q-l_{\max}-b_{\max}(N-1)M}\rfloor
		\end{cases},
	\end{equation}
	the IO characterization in the DD domain can be simplified as
	\begin{equation}
		\label{IO_ZP_ODDM}
		Y[m,n]=\sum_{l=0}^{l_{\max}^{\prime}}\sum_{k=0}^{N-1}X[m-l,(n-k)_{N}]G_{l}(m,k),
	\end{equation} 
	where $X[m,n]\ne0$ holds only for $m_{\min}\leq m\leq m_{\max}$ and $G_{l}(m,k)$ can be derived by \eqref{RCP_DD_G}.
	\begin{IEEEproof}
		The proof is provided in Appendix \ref{th2_ZP_proof}.
	\end{IEEEproof}
\end{theorem}
Considering the delay dispersion caused by the time expansion, the ZP range should also be extended compared with \cite{ZP_ODDM,OTFS_ZP_MRCdetect_TVT}. On the other hand, $l_{\min}^{\prime}$ can be set as $0$ by following the configurations of synchronization in Section \ref{subsec_IO_time}. In this case, $m_{\min}$ can be set as $0$, and ZP is only reserved for the last $M-1-m_{\max}$ multicarrier symbols.\par 
\subsection{Discussions}
\label{subsec_IO_discussion}
From the derivation in this section, it is obvious that even though the physical channel is on-grid, i.e., $l_{i}$ and $k_{i}$ are all integers, the sparse equivalent channel in \cite{ref_ODDM} cannot be acquired due to the DD dispersion caused by the wideband effect, which requires elaborate consideration to assure the accuracy of channel estimation. Meanwhile, even though DSE is ignored and delay resolution is sufficiently high, the off-grid delay is still negligible since the power leakage is proportional to the ratio between the off-grid delay and $T_{s}$, which will not decrease with $T_{s}$ increasing. In fact, a similar conclusion can be deduced for rectangular waveform-based OTFS systems.\par 
On the other hand, even though different equivalent taps may have different affected regions in the time domain, they affect all elements in the DD domain. It indicates that the time domain channel enjoys better sparsity considering DSE, which is more suitable for channel estimation and equalization. Finally, the derived IO characterization can also help improve the sensing accuracy in ODDM-assisted integrated sensing and communication (ISAC) systems, which deserves further investigation in future research. \par 
\section{Numerical Results}
\label{sec_simu}
In this section, numerical results are presented to evaluate the impact of DSE. Two scenarios illustrated in Table \ref{simulation_para_table} are considered, where Type \uppercase\expandafter{\romannumeral1} and \uppercase\expandafter{\romannumeral2} cater to radio-frequency (RF) \cite{ref_ODDM,RCP_ODDM}, and UWA communications \cite{UWA_OTFS,ref_UWAchannel_TSP,ref_UWA_JOE}, respectively. The modeling error ignoring DSE is evaluated by the normalized mean square error (NMSE) defined as $\text{NMSE}=\mathbb{E}\left\{\frac{||\mathbf{H}_{\text{DD}}-\hat{\mathbf{H}}_{\text{DD}}||_{F}^{2}}{||\mathbf{H}_{\text{DD}}||_{F}^{2}}\right\}$, where $\hat{\mathbf{H}}_{\text{DD}}$ denotes the DSE-ignorant DD domain channel matrix. Finally, the speed of each path is generated as $\text{v}_{i}=v_{\max}\cos{\theta_{i}}$, where $\theta_{i}$ is uniformly distributed over $[-\pi,\pi]$. \par 
\begin{table}
	\caption{Simulation Parameters}
	\centering
	\label{simulation_para_table}
	\renewcommand\arraystretch{1.3}
	\begin{tabular}{c|c|c}
		\hline\hline
		Parameters&Type \uppercase\expandafter{\romannumeral1}&Type \uppercase\expandafter{\romannumeral2}\\
		\hline\hline
		Wave speed ($c$)&$3\times10^{8}$ m/s&$1500$ m/s\\
		\hline
		Carrier frequency ($f_{c}$)&5 GHz&12.5 kHz\\
		\hline
		Sampling frequency ($f_{s}$)&15.36 MHz&5 kHz\\
		\hline
		Number of symbols ($M$)&128$\sim$1024&128$\sim$1024\\
		\hline
		Number of subcarriers ($N$)&32, 64&16, 32\\
		\hline
		Roll-off factor&0.1&0.65\\
		\hline
		Delay-power profile&TDL-C~\cite{ref_TDL}&exponential decay~\cite{ref_UWAchannel_TSP}\\		
		\hline\hline
	\end{tabular}
\end{table}
\begin{figure}
	\centering{\includegraphics[width=0.75\linewidth]{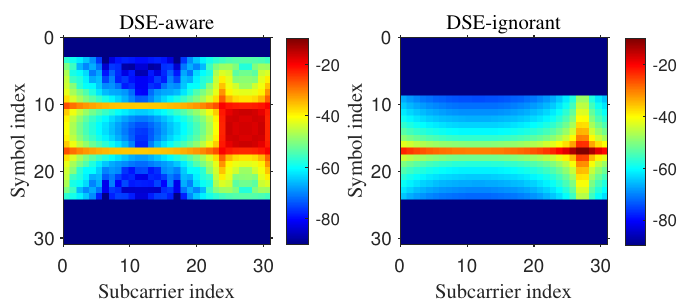}}
	\caption{The received DD domain components (dB) under Type \uppercase\expandafter{\romannumeral2} configurations with $X[m,n]=\delta_{m}\delta_{n}$ under $\tau_{i}-l_{\min}T_{s}=1.5$ ms and $\text{v}_{i}=1$ kn.}
	\label{SimuFig_channel}	
\end{figure}
The difference in DD spread is first presented in Fig. \ref{SimuFig_channel} by plotting the received components in the DD domain with $X[m,n]=\delta_{mn}$ under Type \uppercase\expandafter{\romannumeral2} configurations, where we set $N=32$, $M=128$, $P=1$, $\tau_{i}-l_{\min}T_{s}=1.5$ ms, and $\text{v}_{i}=1$ kn, respectively. Only the effective range in the delay spread is plotted. It is explicit that DSE leads to much more spread in the DD domain due to the interactive dispersion in wideband systems. Meanwhile, the IO pattern with DSE reveals better block sparsity compared with that ignoring DSE, which might be beneficial for channel estimation over off-grid channels.\par 
\begin{figure}[t]
	\centering{\includegraphics[width=0.75\linewidth]{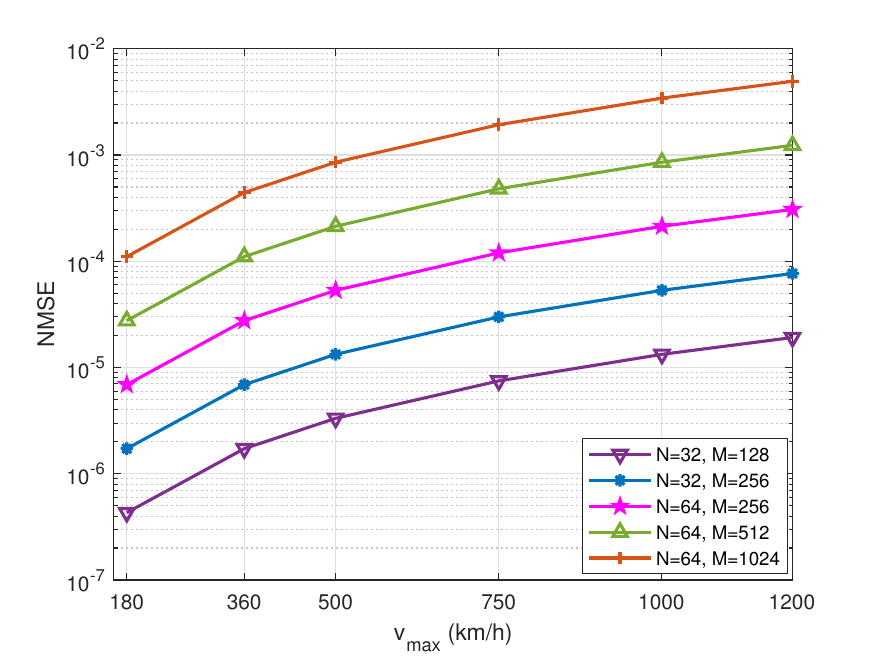}}
	\caption{Modeling NMSE against $v_{\max}$ under Type \uppercase\expandafter{\romannumeral1} channels.}
	\label{SimuFig_NMSE_I}	
\end{figure}
Fig. \ref{SimuFig_NMSE_I} illustrates the modeling NMSE against $v_{\max}$ under Type \uppercase\expandafter{\romannumeral1} channels. The impact of DSE increases with $NM$ and $v_{\max}$ increasing, which confirms our analysis. NMSE of more than $2\times10^{-3}$ occurs when it comes to $(N,M)=(64,1024)$ and the maximum relative mobility of $750$ km/h, which leads to decoding error if DSE is ignored especially in high signal-to-noise (SNR) scenarios and the impact will be more severe when considering the channel estimation \cite{OTFS_DSE_TWC_mine}.\par 
\begin{figure}
	\centering{\includegraphics[width=0.75\linewidth]{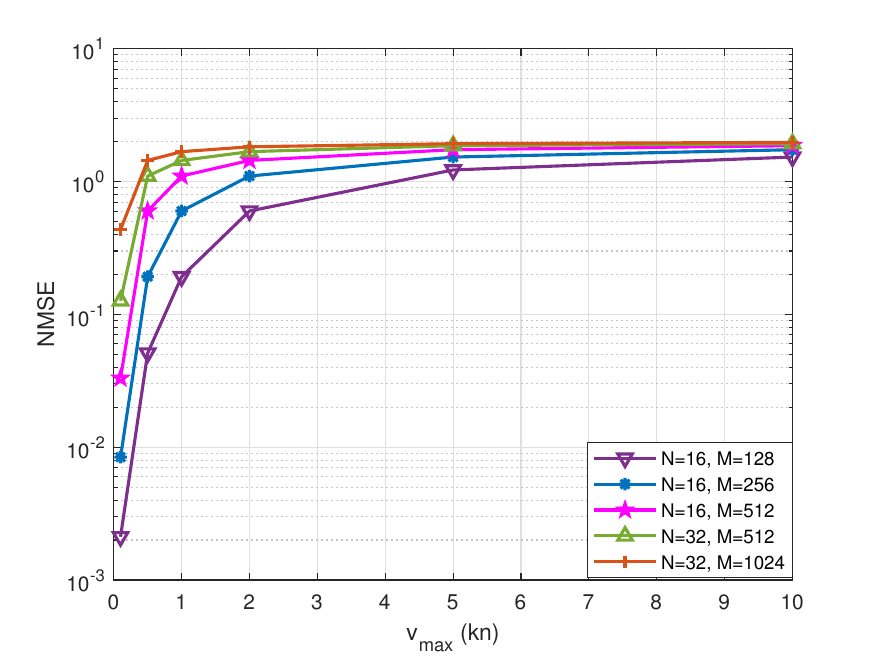}}
	\caption{Modeling NMSE against $v_{\max}$ under Type \uppercase\expandafter{\romannumeral2} channels.}
	\label{SimuFig_NMSE_II}	
\end{figure}
Fig. \ref{SimuFig_NMSE_II} presents the modeling NMSE against $v_{\max}$ under Type \uppercase\expandafter{\romannumeral2} channels and $P=10$. The inter-arrival time of paths is distributed exponentially with the mean value of $1$ ms \cite{ref_UWAchannel_TSP,UWA_OFDM_DL}, which leads to the average delay spread of $10$ ms. The path amplitude follows the Rayleigh distribution with the average power decreasing exponentially with a total decay of $20$ dB for $10$ ms. Even though we have typically small values of $(N,M)$ as $(16,128)$, NMSE will be larger than $100\%$ when $v_{\max}=5$ kn, which caters to the detection failure in \cite{OTFS_DSE_TVT_mine} and indicates the necessity of considering DSE in future research.\par 
\section{Conclusion}
In this paper, we have investigated the IO characterization for ODDM systems with Doppler squint. We first presented the multipath LTV channel model with DSE, where the phenomenon of interactive dispersion in wideband channels is illustrated. The time domain analysis was then established to depict the extra spread more clearly. The characterization was executed for both RCP and ZP scenarios. Finally, numerical results were presented to confirm the significance of considering DSE. The modeling and evaluation in this paper could help better exploit the potential of ODDM modulation by enabling more accurate channel estimation, which is also beneficial for emerging ISAC applications. 
\appendices
\section{Proof of Theorem \ref{th1_RCP_ODDM_IO}}
\label{th1_RCP_ODDM_IO_proof}
\begin{figure*}
	\begin{equation}
		\small
		\label{RCP_ODDM_derive}
		\begin{aligned}
			Y[m,n]&=\frac{1}{\sqrt{N}}\sum_{\dot{n}=0}^{N-1}y[m,\dot{n}]e^{-j2\pi\frac{n\dot{n}}{N}}\\
			&=\frac{1}{\sqrt{N}}\sum_{\dot{n}=0}^{N-1}\left(\sum_{l=0}^{m}x[(m-l)_{M},\dot{n}]g_{l}(m,\dot{n})+\sum_{l=m+1}^{l_{\max}^{\prime}}x[(m-l)_{M},(\dot{n}-1)_{N}]g_{l}(m,\dot{n})\right)e^{-j2\pi\frac{n\dot{n}}{N}}\\
			&=\frac{1}{N}\sum_{\dot{n}=0}^{N-1}\left(\sum_{l=0}^{m}\sum_{n^{\prime}=0}^{N-1}X[(m-l)_{M},n^{\prime}]g_{l}(m,\dot{n})+\sum_{l=m+1}^{l_{\max}^{\prime}}\sum_{n^{\prime}=0}^{N-1}X[(m-l)_{M},n^{\prime}]e^{-j2\pi\frac{n^{\prime}}{N}}g_{l}(m,\dot{n})\right)e^{-j2\pi\frac{n\dot{n}}{N}}e^{j2\pi\frac{n^{\prime}\dot{n}}{N}}\\
			&\overset{(a)}{=}\frac{1}{N}\sum_{\dot{n}=0}^{N-1}\sum_{l=0}^{l_{\max}^{\prime}}\sum_{k=0}^{N-1}X[(m-l)_{M},(n-k)_{N}]\phi[m-l,n-k]g_{l}(m,\dot{n})e^{-j2\pi\frac{\dot{n}k}{N}}\\
			&=\sum_{l=0}^{l_{\max}^{\prime}}\sum_{k=0}^{N-1}X[(m-l)_{M},(n-k)_{N}]\phi[m-l,n-k]\times\frac{1}{N}\sum_{\dot{n}=0}^{N-1}\left(\sum_{i=1}^{P}h_{i}g_{l}^{i}(m,\dot{n})\right)e^{-j2\pi\frac{\dot{n}k}{N}}\\
			&=\sum_{l=0}^{l_{\max}^{\prime}}\sum_{k=0}^{N-1}X[(m-l)_{M},(n-k)_{N}]\phi[m-l,n-k]\sum_{i=1}^{P}h_{i}\times \underbrace{e^{j2\pi\frac{k_{i}m}{NM}}\frac{1}{N}\sum_{\dot{n}=0}^{N-1}e^{-j2\pi\frac{\dot{n}(k-k_{i})}{N}}a\Big(\big(l-l_{i}+b_{i}(m+\dot{n}M)\big)T_{s}\Big)}_{G^{i}_{l}(m,k)}
		\end{aligned}
	\end{equation}
	\hrulefill	
\end{figure*}
By substituting \eqref{IFFT_ODDMTx} and \eqref{FFT_receiver} into \eqref{time_IO_tap}, \eqref{RCP_ODDM_derive} can be derived at the top of the next page, where $(a)$ is deduced by changing the variable as $k=(n-n^{\prime})_{N}$. The proof of \textbf{Theorem \ref{th1_RCP_ODDM_IO}} can then be completed by employing the final conclusion in \eqref{RCP_ODDM_derive}.\par
\section{Proof of Theorem \ref{th2_ZP_ODDM}}
\label{th2_ZP_proof}
The configurations of ZP is actually set to satisfy no ISI in the time domain, i.e., $\psi^{i}_{m,m^{\prime},\dot{n},\dot{n}^{\prime}}=0$ for $\forall \dot{n}\ne\dot{n}^{\prime}$ in \eqref{time_base_ini}. Bear in mind that $a(t)=0$ holds for $|t|\geq QT_{s}$ with $2Q\ll M$. When we have $\dot{n}>\dot{n}^{\prime}$, it should be guaranteed that
\begin{equation}
	m+\dot{n}M-m^{\prime}-\dot{n}^{\prime}M-l_{i}+b_{i}(m+\dot{n}M)\geq{Q}
\end{equation} 
for $\forall~0\leq{m}\leq{M-1}$. Therefore, we can derive that
\begin{equation}
	\label{ZP_condition_right}
	\small
	\begin{aligned}
		m^{\prime}&\leq\mathop{\min}\Big((\dot{n}-\dot{n}^{\prime})M+m-Q-l_{i}+b_{i}(m+\dot{n}M)\Big)\\
		&=M-Q-l_{\max}-b_{\max}(N-1)M,
	\end{aligned}
\end{equation}
where we assume $|b_{\max}|<1$. On the other hand, when we have $\dot{n}<\dot{n}^{\prime}$, the following equation should be satisfied as
\begin{equation}
	m+\dot{n}M-m^{\prime}-\dot{n}^{\prime}M-l_{i}+b_{i}(m+\dot{n}M)\leq{-Q}.
\end{equation}
As a result, it can be deduced that
\begin{equation}
	\label{ZP_condition_left}
	\small
	\begin{aligned}
		m^{\prime}&\geq\max\Big((\dot{n}-\dot{n}^{\prime})M+m+Q-l_{i}+b_{i}(m+\dot{n}M)\Big)\\
		&=Q-l_{\min}-1+b_{\max}\Big((N-1)M-1\Big)
	\end{aligned}
\end{equation}
by setting $m=M-1$ and $\dot{n}=N-2$. Based on the deduction in \eqref{ZP_condition_right} and \eqref{ZP_condition_left}, to satisfy no ISI in the time domain, the range of ZP can be set as \eqref{ZP_range_eq}. \eqref{time_IO_tap} can then be simplified as
\begin{equation}
	y[m,\dot{n}]=\sum_{l=0}^{l_{\max}^{\prime}}x[m-l,\dot{n}]g_{l}(m,\dot{n}),
\end{equation}
where $l_{\min}^{\prime}$ is assumed to be $0$ by enabling appropriate synchronization as illustrated in Section \ref{subsec_IO_time}. Then the deduction in \eqref{RCP_ODDM_derive} can be repeated by ignoring the ISI term to derive \eqref{IO_ZP_ODDM}, which completes the proof of \textbf{Theorem \ref{th2_ZP_ODDM}}.
\section*{Acknowledgment}
This work was supported in part by the National Natural Science Foundation of China under Grants 624B2079 and 62271284.

\bibliographystyle{IEEEtran}
\bibliography{ref-sum}

\end{document}